\documentclass[11pt]{article}
\usepackage{authblk}

\begin{document}

\title{Report of the Topical Group on Solid State Detectors and Tracking for Snowmass 2021}


\author{Anthony Affolder, Artur Apresyan, Steven Worm \\ {\it for the Snowmass Instrumentation Frontier Solid State Detector and Tracking community:} \\   M.~Albrow, D.~Ally, D.~Ambrose, E.~Anderssen, N.~Apadula, P.~Asenov, W.~Armstrong, M.~Artuso, A.~Barbier, P.~Barletta
% ...
}

\maketitle

\section{Abstract}
    Tracking detectors are of vital importance for collider-based high energy physics (HEP) experiments. The primary purpose of tracking detectors is the precise reconstruction of charged particle trajectories and the reconstruction of secondary vertices. The performance requirements from the community posed by the future collider experiments require an evolution of tracking systems, necessitating the development of new techniques, materials and technologies in order to fully exploit their physics potential. In this article we summarize the discussions and conclusions of the 2022 Snowmass Instrumentation Frontier subgroup on Solid State and Tracking Detectors (Snowmass IF03).

\section{Executive Summary}

Tracking detectors are of vital importance for collider-based high energy physics (HEP) experiments. The primary purpose of tracking detectors is the precise reconstruction of charged particle trajectories and the reconstruction of secondary vertices. The performance requirements from the community posed by the future collider experiments require an evolution of tracking systems, necessitating the development of new techniques, materials and technologies in order to fully exploit their physics potential. 

Relative to the currently operating systems and their upgrades, the technical requirements for tracking detectors (trackers) in the next 20-40 years are significantly more stringent, such as tolerances to fluences 1-2 orders of magnitude higher, larger areas at lower costs, segmentation and position resolution 2-4 times finer and precision timing resolution, radiation length per layer from 0.1-1\% $\rm{X_{0}}$ and integration of novel radiation-hard materials. 


Technological developments currently underway aim to address these issues, and the successful completion of the programs outlined below 
requires focused efforts from the community on the steady development and refinement of existing technologies, and the pursuit of novel 
``blue sky'' technologies to enable transformative progress. The HEP community gathered at Seattle Snowmass Summer Meeting in 2022 identified the following key directions for the near-term priorities of the solid-state tracking:

\begin{itemize}
    \item \textbf{IF03-1}: Develop high spatial resolution pixel detectors with precise per-pixel time resolution to resolve individual interactions in high-collision-density environments
    \item \textbf{IF03-2}: Adapt new materials and fabrication/integration techniques for particle tracking in harsh environments, including sensors, support structures and cooling
    \item \textbf{IF03-3}: Realize scalable, irreducible-mass trackers in extreme conditions
    \item \textbf{IF03-4}: Progress advanced modeling for simulation tools, developing required extensions for new devices, to drive device design.  
    \item \textbf{IF03-5}: Provide training and workforce maintenance to enable future tracking systems to be designed, developed, constructed and simulated.
    \item \textbf{IF03-6}: Nurture collaborative networks, provide technology benchmarks and roadmaps and funding in order to develop required technologies on necessary time scales, costs and scope.
\end{itemize}

Eight white papers on solid state trackers were submitted during the 2022 Snowmass~\cite{arxiv.2202.11828,arxiv.2203.06093,arxiv.2203.06773,arxiv.2203.07626,arxiv.2203.08554,arxiv.2203.13900,arxiv.2203.06216,arxiv.2203.14347}, focusing on a broad set of R\&D topics that the US community is actively pursuing. While these papers don't completely capture the full breadth and depth of the R\&D carried out in the US, they highlight the key challenges that future trackers will face and summarize the main research thrusts actively pursued by the community. 

Throughout the tracker community, many new, novel tracking technologies are being investigated.
Silicon pixel detectors based on Monolithic Active Pixels (MAPS) are, for example, being developed for a wide range of applications.  Recent devices show significant improvements in pixel granularity, readout speed and radiation tolerance.  Also addressing detector speed and radiation performance are so-called 3D sensors, where the collection electrodes are orthogonal to the plane of the detector.  Developed in both silicon and diamond subtrates, 3D sensors serve as a first step into wide-bandgap semiconductor materials.  Next, thin film detectors offer a radical new way to manufacture large-area and low-cost tracking detectors.  Finally 
quantum-enabled detection is being investigated for trackers in novel architectures and materials, with the promise of achieving extremely fast response and high granularity with only a minimal detector thickness.

Advanced 4-dimensional trackers with ultra fast timing (10-30~ps) and extreme spatial resolution (O(few~$\mu$m)) represent a new avenue in the development of silicon trackers, enabling a variety of physics studies which would remain out of experimental reach with the existing technologies. Several technology solutions are being currently pursued by the community to address the challenges posed by various experiments, both for the sensors and the associated electronics. Bringing in technological innovation and fully exploiting the potential of future detectors through the usage of ultra fast timing is a unique and exciting  opportunity for the particle physics community. In order to reach this goal, it is of paramount importance in the coming years to undertake thoroughly the R\&D studies to investigate how to best combine timing with spatial information.

The packaging and integration of the sensors and readout electronics will become more critical for future experiments, as device segmentation decreases to mitigate the increased track density. Bump bonding technologies have nearly reached their limits; more advanced 3D packing technologies including wafer-to-wafer and die-to-wafer hybrid bonding and through silicon vias (TSV) have the potential to meet the goals of future particle physics experiments.   Collaboration between research groups and industrial partners with this expertise is crucial to introduce this technology to the HEP community.  This research will take consider time and efforts to have confidence to use these technique in future large-scale tracking systems, and as such need to being urgently.  

Detector mechanics will also play a significant role in future detector's performance.  Material necessary for cooling and structural stability will be the lower bound on the radiation length for future tracking systems.   Increased segmentation will lead naturally to larger power densities; in order to minimize material, solutions with integrated services and cooling are necessary.   A holistic approach to design, simulation and manufacturing will be required. Novel materials, new cooling and composite manufacturing technique will need to be developed in order to reach the targeted performance.   

To develop these new technologies, simulations of the properties of silicon and novel sensor materials throughout the lifetime of the experiments will be critical.  These studies can drive device design included implant locations, size and strength to the most promising directions of development.   To reach their full potential, further developments are needed to improve accuracy, precision and new devices.   With this research, the performance of future experiments can be better predicted for its full life cycle prior to construction.

\section{Requirements}

Tracking detectors are key to address science challenges and their development is based on our understanding of fundamental laws of physics.
Therefore, there is a “virtuous cycle” between fundamental laws of nature and enabling detector concepts and techniques. Such virtual cycle must remain strong and unbroken and in turn leads to a greater physics discoveries.

At the Energy Frontier, there are two major drivers for detector R$\&$D: detectors for future hadron and muon colliders and development of advanced technologies for the future ${e^+e^-}$ machines. 
The challenge of the most intense tracking regions at future hadron colliders will be addressed by hybrid pixel detectors, and the breakthrough technology is expected to come from CMOS sensor developments, which are currently being pushed to unprecedented levels of radiation hardness and rate capabilities. For future applications in particle physics, alternative technologies, by employing beyond state-of-the-art interconnection technologies, such as 3D vertical integration, through-silicon-vias (TSV), or micro bump-bonding, which, while retaining the advantages of separate and optimized fabrication processes for sensor and electronics, would allow fine pitch interconnect of multiple chips.
The Basic Research Needs for High Energy Physics Detector Research \& Development document~\cite{osti_1659761} compiles a list of requirements for trackers
at the next generation of energy frontier experiments focused on precision Higgs and SM physics, and searches for BSM phenomena.

Emerging novel vertex and tracking detector technologies are the vital backbone for the success at a future electron-positron machines. These will operate in an environment with high (continuous or bunched) beam currents, a minimum distance from beam axis of about 20 mm, a requirement of $< 5~\mu m$ single point resolution, high granularity ($< 30 \times 30~\mu m^2$), power dissipation ($<50~mW/cm^2$), low mass ($\sim 0.1~\%$ of $X_0$, or 100 $\mu m$ Si-equivalent per layer). 
The clear challenge is unprecedented spatial resolution, to be achieved with ultra-small pixels, and thus extremely low material budget. Very thin detector assemblies are mandatory, while providing high stiffness and stability. 
The tracking resolutions should enable high-precision reconstruction of the recoil mass in the $e^+e^- \rightarrow Zh$ process, as shown in Table~\ref{tab:physics-req}, and allow very efficient $b$ and $c$ tagging and $\tau$-lepton identification through the reconstruction of secondary vertices.
 The detector solenoid magnetic field must be limited to 2 T 
 for the Tera-Z operation at future circular electron-positron colliders, to avoid a blow up of the vertical beam emittance and a resulting loss of luminosity. The 2 T magnetic field limit is not a significant problem since the momentum scale of the produced partons is typically distributed around 50 GeV and does not exceed 182.5 GeV. 
In general, the future demands at electron-positron Higgs Factories for high resolution (granularity) and low material budget on one hand, and low power consumption on the other hand, exceed significantly what is the state-of-the-art today.

\begin{table}[!ht]
\begin{center}
 \caption{Physics goals and detector requirements~\cite{osti_1659761,MuonCollider:2022ded}.}
\begin{tabular}[c]{|c|c|c|l|}
\hline
Initial state	&	Physics goal	& Detector & 	Requirement	\\
\hline
\hline
$e^+e^-$ & $h$ZZ sub-\% & Tracker & $\sigma_{p_T}/p_{T}$=0.2\% for $p_T <$ 100 GeV \\ 
&&& $\sigma_{p_T}/p_{T}^2=2 \cdot 10^{-5}/$ GeV for $p_T >$ 100 GeV \\ 
& $ hb\bar{b}/h c\bar{c}$ & Tracker & $\sigma_{r\phi} = 5\oplus 15 ( p\sin\theta^{\frac{3}{2}})^{-1}\mu$m\\ 
&&& 5$~\mu$m single hit resolution \\ 
&&& per track timing resolution of 10 ps\\ 
\hline
pp-100 TeV  & Higgs \& BSM & Tracker & $\sigma_{p_T}/p_{T}$=0.5\% for $p_T <$ 100 GeV \\ 
&&& $\sigma_{p_T}/p_{T}^2=2 \cdot 10^{-5}/$ GeV for $p_T >$ 100 GeV \\ 
&&& $ 300 $ MGy and $\approx$ 10$^{18}$  n$_{eq}$/cm$^{2}$\\ 
&&& per track timing resolution of 5 ps \\
\hline
$\mu^+\mu^-$  & Higgs \& BSM & Tracker 
& $\sigma_{p_T}/p_{T}$=0.2--0.3\% for $p_T <$ 100 GeV \\
&&& $\sigma_{p_T}/p_{T}^2=2 \cdot 10^{-5}/$ GeV for $p_T >$ 100 GeV \\
&&& 100 kGy and $\approx$ 10$^{15}$  n$_{eq}$/cm$^{2}$\\ 
&&& 0.01 rad angular resolution \\
&&& 5$~\mu$m single hit resolution\\ 
&&& 30 ps timing resolution\\ 
\hline
\end{tabular}
\label{tab:physics-req}
\end{center}
\end{table}

Picosecond timing information will be the vital backbone for the success of future hadron colliders, driven by the general push towards higher luminosities at the next generation of colliders. The addition of the timing information will allow one to implement 4D reconstruction of the primary collision vertex - joint pattern recognition, based on timing and tracking detectors.
In particular, the required accuracy in the ATLAS and CMS at HL-LHC will be dictated by the pile-up induced backgrounds and the nominal interaction distribution within a single bunch crossing ($\sigma_z \sim$~5 cm, $\sigma_t \sim$~170 ps rms). 
Furthermore, a factor-of-five larger pileup at a 100 TeV hadron collider than at the HL-LHC will pose stringent requirements on the detector
design~\cite{Aleksa:2019pvl}: future hadron colliders require timing information of the order of 5-10 ps per track to suppress pileup and correctly assign tracks to vertices.
At a multi-TeV muon collider to reject a significant fraction of beam induced background, accurate timing information with a resolution of 30 ps is assumed to be available in the vertex detectors~\cite{MuonCollider:2022ded}.
Aiming for an excellent position and timing resolution ($\sim$~10~ps and $\sim$~10~$\mu m$) with GHz counting capabilities to perform 4D tracking, LGAD sensors 
represent a very attractive option for PID and TOF applications.

The 4D-tracking systems will play a crucial role in LLP exploration allowing to efficiently reconstruct position-time vertice, including those with large displacement, and reduce backgrounds including beam-induced contributions.
4D-trackers will also enable complementary TOF and dE/dx measurement in the same sensor critical for PID
and heavy stable charged particle (HSCP) searches. In the latter case, the TOF measurement yields complementary information of
potential evidence of new HSCP's, as recently reported by ATLAS~\cite{ATLAS:2022pib},
and signal characterization in case of discovery.
An ultimate concept is to develop 4D real-time tracking system for a fast trigger decision and to exploit 5D imaging reconstruction approach, if space-point, picosecond-time and energy information are available at each point along the track.

At future hadron colliders, detector capabilities to reconstruct highly boosted objects are fairly challenging (for instance, the average $Z$ boson from $ZZ$ production would shower mostly within a single LHC calorimeter cell).   This challenge is accentuated by so-called ``hyper-boosted'' jets, whose decay products are collimated into areas the size of single calorimeter cells. Holistic detector designs that integrate tracking, timing, and energy measurements are needed to mitigate for these conditions~\cite{larkoski2015tracking,trackbased2013,track-assistedmass2019, trackingspannowsky2015,Gouskos:2642475}.

Future hadron and muon collider experiments at the Energy Frontier have an additional requirement of radiation hardness of tracking systems.
Despite their quite complicate fabrication process, that can be both double- and single-sided, 3D detectors are the most radiation-hard technology to-date demonstrating good performance even up to a fluence of $3 \times 10^{16}$ $neq/cm^2$ and time resolution of 30 ps. New developments are being pursued to address the challenges of radiation hardness, small pitch with narrower columns, thinner devices, better breakdown behaviour, improved on-wafer selection and lower costs.

The Rare Processes and Precision Measurements (RPF) frontier encompasses several different experimental programs who share the common aim of searching for new physics through its manifestation in rare and forbidden decays in the SM, as well as in precise measurements sensitive to subtle deviations from SM expectations. Four main categories of experiments were considered: heavy flavor, lepton flavor violation, electric dipole moment, and the dark sector. 
This represents a data-driven approach to search for signatures that may lead to some understanding of the key feature of the new physics, namely its mass scale and coupling to SM particles. This approach has been very fruitful in the past. 

The continual exploration of new physics in heavy flavor is pursued by LHCb at the LHC hadron collider\cite{lhcb-whitepaper}, with a brand new detector being commissioned right now, and an ambitious upgrade planned for the HL-LHC. In this phase, the detector operates at a  peak luminosity pf $1.7\times 10^{34}{\rm cm^{-2} s^{-1}}$ and a pile-up of $\sim 40$. Belle II\cite{belle2upgrade-whitepaper} is also consider upgrades. The first goal is to reach an integrated luminosity of 50 ab$^{-1}$ by 2035. Detector upgrades are considered, in the second long shutdown possibilities include a CMOS vertex detector. Longer terms upgrades of the luminosity are considered. Lighter quark flavor physics include rare $K$ decays,  $\eta-\eta^\prime$ factories\cite{redtop-whitepaper}, and rare $\pi$ decays. 

Charged lepton flavor violation is another pathway to uncover indirect evidence for new physics. Example experiments that will study rare $\mu$ decays in the next few decades are MEG II, COMET and Mu3e (off-shore), and Mu2e and its upgrade Mu2eII, taking data at Fermilab. If we consider experiments sensitive to $\mu$ conversion in the field of a nucleus, the signature of a signal event is an electron with an energy near the rest mass of the muon, characteristic of neutrinoless $\mu\to\ e$ conversion in the field of a nucleus. Separating the signal from the background from cosmic ray processes requires low mass tracking with excellent momentum resolution. This is currently accomplished with an extremely low mass gaseous detector. The Mu3e experiment requires excellent momentum, position and tracking information to achieve the desired 10$^{-16}$ branching ratio sensitivity to $\mu^+\to e^*e^-e^+$\cite{Hesketh:2022wgw}.
 To accomplish this, a low-mass HV-CMOS tracking detector is an integral component. 

Some of the experiments considered rely on state of the art solid state detector systems to achieve their desired sensitivities. In the following we will discuss the requirements of this broad experimental physics program with respect to four main technological aspects: time stamp specifications on single hits and tracks, spatial resolution and granularity, environmental conditions such as radiation tolerance, and system aspects such as scalability, material budget and cooling requirements.

A precision time stamp, of the order of tens of ps, makes it possible to achieve a   4D tracking reconstruction algorithm, with the time dimension fully integrated as a fourth dimension.  The Vertex Locator (VELO) for the LHCb Upgrade2  requires  with 20~ps time resolution per track will maintain high primary vertex finding efficiency at higher $pp$ collision density while also providing a precision time stamp to each collision~\cite{lhcb-whitepaper}. Similar time stamps on other tracking subsystems, and particle identification devices (calorimeter, hadron identification systems), enhances the speed of  the all-software trigger, and enables the use of time of flight for particle identification. Dedicated timing layers in the vertex detector and in the mid-section of the calorimeter, close to the point where the maximum shower is developed, are also studied. Their hit resolution should be of the order of 25 ns or better. The VELO must do this while maintaining a high spatial resolution at the 10~$\mu$m level. The requirements for a timing layer integrated in the VELO telescope is a 25 ns time stamp at the hit level. An option considered in this upgrade is a timing layer in the calorimeter region of maximum shower development with similar specifications. An upgrade of the NA62 kaon experiment also demonstrates the benefits of 4D tracking of the incoming beam with sub-50~ps resolution, and then using time measurements in final state detectors to match decay products to incoming particles at a rate of 3 ${\rm GHz}$~\cite{kaons-whitepaper}. Other experiments may require timing resolutions at the sub-100~ps level, as well.

High spatial granularity while maintaining a low material budget is the most desirable feature in several applications. 
For example, the MuPix pixel sensors implemented for the M3e experiment are being studied as a possible starting point for the design of the Upstream Tracker and Mighty Tracker for the LHC upgrade~\cite{lhcb-whitepaper}.  Here a pixel size of $50\times 150~\mu m^2$ is considered. 

Experiments in RF also need development of tracking detectors with large areas and low material budgets. Tracking in the proposed REDTOP experiment requires radiation lengths per layer $\sim 0.1\%$ $\rm{X_{0}}$ if implemented using silicon sensors~\cite{redtop-whitepaper}. Proposed future upgrades of the Belle 2 vertex detector should have a radiation length per layer of 0.2-0.7\% $\rm{X_{0}}$~\cite{belle2upgrade-whitepaper}. 

The radiation environment is also a driver in the technology choices, especially in applications involving hadron beams. For example, in LHCb upgrade 2, the VELO detector  needs to operate at fluences up to the $10^{16}~n_{\rm eq}/{\rm cm}^2$ level, while the upstream tracker needs to withstand fluences up to $\sim 3\times10^{15}~n_{\rm eq}/{\rm cm}^2$. 

\section{Silicon Radiation Detectors for High Energy Physics Experiments}


There are currently a variety of  tools available for simulating the properties of silicon sensors before and after irradiation.  These tools include finite element methods for device properties, dedicated annealing models, and testbeam/full detector system models.  No one model can describe all of the necessary physics.  Most of these models are either fully or partially developed by HEP scientists and while there are many open-source tools, the most precise device property simulations rely on expensive, proprietary software. 

The development of these simulations happens inside existing experimental collaborations and within the RD50 Collaboration at CERN.  RD50 is essential for model research and development and provides an important forum for inter-collaboration exchange.

While existing approaches are able to describe many aspects of signal formation in silicon devices, even after irradiation and annealing, there is significant research and development (R\&D) required to improve the accuracy and precision of these models and to be able to handle new devices (e.g. for timing) and the extreme fluences of future colliders.  The US particle physics community can play a key role in this R\&D program, but it will require resources for training, software, testbeam, and personnel. 


For example, there is a great need for (1) a unified microscopic model of sensor charge collection, radiation damage, and annealing (no model can currently do all three), (2) radiation damage models (for leakage current, depletion voltage, charge collection) with uncertainties (and a database of such models), and (3) a measurement program to determine damage factors and uncertainties for particle types and energies relevant for current and future colliders.

This program of work is critical for the future success of collider physics.   With these tools, the performance of future experiments would be better predicted throughout their lifetimes.   The simulations will also be critical in directing the R\& D for the next generation of complex sensors into the most promising avenues to realize the necessary performance to reach our ultimate physics goals.  

\section{Novel Sensors for Particle Tracking}

The proposed future collider experiments pose unprecedented challenges for event reconstruction, which require development of new approaches in detector designs. Research in particle tracking detectors for high energy physics application is underway with a goal of improving radiation hardness, achieving improved position vertex and timing resolution, simplifying integration, and optimizing power, cost, and material. Several technologies are currently being investigated to develop technologies for the future, which approach these goals in complementary ways. 

Silicon sensors with ``3D technology" have electrodes oriented perpendicular to the plane of the sensor.
These show promise for compensation of lost signal in high radiation environments and for separation of pileup events by precision timing.  New 3D geometries involving p-type trench electrodes spanning the entire thickness 
of the
detector, separated by lines of segmented n-type electrodes for readout, promise improved uniformity, timing resolution, and radiation resistance relative to established devices operating effectively at the LHC.  The 3D technology is also being realized in diamond substrates, where column-like electrodes may be placed inside the detector material by use of a 130 fs laser with wavelength 800~nm. 
Present research
aims for operation with adequate signal-to-noise ratio at fluences approaching $10^{18}~n_{\rm eq}/{\rm cm}^2$, with timing resolution on the order of 10 ps. 

Monolithic Active Pixel Sensors (MAPS), in which charge collection and readout circuitry are combined in the same pixel, have been shown to be a promising technology for high-granularity and low material budget detector systems. MAPS have several advantages over traditional hybrid pixel detector technologies, as they can be inexpensively fabricated in standard CMOS imaging processes, back-thinned or back-illuminated, have demonstrated high radiation hardness, and can be stitched or tiled for large, low-mass arrays.
MAPS offer the possibility of individual pixel readout at MHz speeds, with low power consumption and powered from a low voltage supply. 
 
 Many MAPS pixel geometries have been explored, but the close connection of a sensor and front-end amplifier, without the need for external interconnections, holds the promise of reducing the input capacitance significantly, and hence extremely low-noise designs are possible.
The reduction of the noise floor means that even small signals, for example from thinned low-power sensors, can yield satisfactorily high S/N. 
The next generation of MAPS will target improved performance in speed, resolution, and radiation tolerance, and also incorporate developments in the systems integration that are needed for large scale use at a reasonable cost. Prototype detectors are currently being designed to addresses the integration issues associated with making large arrays, by first developing an intermediate scale array of tens of m$^{2}$ which include full powering and readout of large numbers of MAPS. 
Many different MAPS detector architectures are in development, and are being adapted to a diverse set of applications and environments; from hadron colliders to the EIC and ILC, to space-based tracker applications in $\gamma$-ray satellites.

Thin film detectors  have the potential to be fully integrated,
while achieving large area coverage and low power consumption with low dead material and low cost.  Thin flim transistor technology uses crystalline growth techniques to layer materials, such that monolithic detectors may be fabricated by combining layers of thin film detection material with layers of amplification electronics using vertical integration. 

The DoTPiX pixel architecture has been proposed on the principle of a single n-channel MOS transistor, in which a buried quantum well gate performs two functions: as a hole-collecting electrode and as a channel current modulation gate.  The quantum well gate is made with a germanium
silicon substrate. The active layers are of the order of 5$~\mu m$ below the surface, permitting detection of minimum ionizing particles.  This technology is intended to achieve extremely small pitch size to enable trigger-free operation without multiple hits in a future linear collider, as well as simplified reconstruction of tracks with low transverse momentum near the interaction point.

Lastly, a technology is under development in which a novel ultra-fast scintillating material employs a semiconductor stopping medium with embedded quantum dots.  The candidate material, demonstrating very high light yield and fast emission, is a GaAs matrix with InAs quantum dots. The first prototype detectors have been produced, and pending research goals include demonstration of detection performance with minimum ionizing particles, corresponding to signals of about 4000 electron-hole pairs in a detector of 20$~\mu m$ thickness.  A compatible electronics
solution must also be developed.  While the radiation tolerance of the device is not yet known, generally quantum dot media are among the most radiation hard semiconductor materials.

\section{4-Dimensional trackers}
\label{sec:if03_4Dtrackers}

Future collider experiments call for development of tracking detectops with 10-30~ps timing resolution, in addition to excellent position resolution, i.e. 4-Dimensional trackers. Time resolution of the future 4D tracker detector can be factorized into contributions from the sensor itself, and those from the readout electronics. The overall detector system should contain a sensor with short drift time, high signal to noise, limited thickness in the path of a MIP to reduce the Landau fluctuations, and small TDC bin size. Several technologies to address these needs are being developed and are introduced in this section.

One approach to meet these requirements utilizes Low Gain Avalanche Detectors (LGADs), and adapts them to achieve the fine segmentation necessary for tracking. The main limitation of the DC-coupled LGADs is the presence of JTE to interrupt the gain layer, which introduces inactive regions on the sensors. Several modifications to LGAD technology have been proposed and demonstrated to make LGAD sensors suitable for tracking sensors with 100\% fill-factor, achieving excellent timing and position resolution. The most advanced high granularity designs realized so far are the AC-coupled LGAD (AC-LGAD). Because the gain layer is uninterrupted, the electrodes can be designed with smaller pitch and size than standard LGADs. A key feature of AC-LGADs is the signal sharing between electrodes, which can be used to obtain the spatial resolution necessary for future colliders with a reduced number of readout channels, achieving simultaneous 30~ps and 5~$\mu$m resolutions. Another design geared towards 100\% fill factor are the ``Deep-Junction'' (DJ-LGAD), which are formed by abutting thin, highly-doped p$+$ and n$+$ layers. 
Trench-isolated (TI) LGADs address the fill-factor limitation of LGADs by substituting the JTE with narrow trenches filled with dielectric material. The width of the trenches is about 1~$\mu$m, allowing for a substantial reduction of the distance between gain layers of neighboring channels with respect to standard LGADs. Intense R\&D is ongoing to improve the radiation tolerance of LGADs beyond $10^{16}~n_{\rm eq}/{\rm cm}^2$, by e.g. exploiting the observation that deep and narrow gain layers are more radiation hard. Such ``Buried Layer LGADs'' achieve the deep gain layer by implanting the boron layer at low energy and then to burying it under a few microns of epitaxially grown silicon.

Another thrust to develop 4D-tracking sensors is through a usage of sensors with 3D geometries, or closer integration with electronics via monolithic pixel sensors, or adoption of new materials in sensor design. Since charge carriers are collected perpendicularly to the sensor thickness in sensors with 3D-geometries, the time uncertainties due to nonuniform ionisation density, delta rays and charge carrier diffusion are minimized. Prototypes have demonstrated promising results with time resolution below 20~ps. New materials, such as SiC detectors, can operate at very high temperatures, and are known to exhibit significant radiation hardness and extremely low leakage currents. Although the adoption of this technology in HEP has been slow, recent advances in wafer size and quality are enabling compatibility with processing tools developed for silicon devices, and allowing the price per device to sharply decrease. 

Going even beyond 4D-sensors are designs that aim to simultaneously measure not only the position and time, but also the angle of passing tracks. These kind of sensors would dramatically reduce the complexity of detector modules, and enable unprecedented reconstruction capabilities on the front-end. Double Sided LGADs (DS-LGADs) achieve this goals by adding a readout layer to the p-side of the LGAD structure, which allows one to also measure signals from the slower-drifting holes. The signal p-side collects two components, holes from the primary ionization followed by the larger number of holes generated at the gain layer. This provides a unique signature of the pattern of charge deposit within the device that could enable measurement of the track angle as well.  Another approach to achieve the same goal is through detection of the Shockley-Ramo current induced at the readout electrode from mobile charge carriers within the sensor. This current has a very fast rising edge ($\sim$15~ps) and can be used to precisely timestamp track hits, and is a promising direction for establishing 5D-sensor designs.

The timing ASICs under development for the ATLAS and CMS timing upgrades, named ALTIROC and ETROC, represent revolutionary steps forward as the first readout chips to bring O(10~ps) timing to collider experiments. However, they are able to use significantly more space and power than high density ASICs designed for trackers with fine pitch and limited material. Compared to RD53A, the timing chips use several hundred times more power and area per channel. The primary challenges to transform them into chips for 4D tracking will be to minimize both the power consumption and the channel size. The high luminosity of future hadron colliders will require trackers capable to survive in extreme radiation environments (accumulating a dose of up to 30~GRad and $10^{18}$ neutrons/cm$^2$). Currently there are  several projects with the aim to make advance in these areas. Efforts are geared towards a specific application, however the goals are common: bandwidth optimization, low noise, low area, time resolution and power dissipation. The CERN’s EP R\&D WP5 chip aims to reduce the total area of the circuitry by moving to 28~nm node with the design of the low-power and compact TDC, while the Silicon-Germanium based ASIC developed by Anadyne in TowerJazz 130~nm is expected to have 0.5 mW per channel (front end and discriminator) with 10 ps of timing resolution for 5 fC charge. The CFD-chip ASIC developed at FNAL in 65~nm node was demonstrated to deliver $\sim$10 ps jitter for 15 fC of injected charge, using a constant-fraction-discriminator implementation.

\section{Integration}


In the past years, HEP experiments have been mostly relying on bump bonding for high density pixel sensor to ASIC connection. 
The bump bonding technology was proven to be reliable; however, it is known to have several limitations: it only works down to 20-50~$\mu m$ of pitch and has yield issues for finer connections. 
Furthermore, the solder balls used for the connection increase the input capacitance to the amplifier and hence the noise. 
With bump bonding,the sensor or chip need side extensions to have external connections.
In terms of mechanical properties the resulting connection is subject to heat stress since it involves different materials; the minimum thickness is also limited since both chip and sensor need a thick enough support wafers in order to meet alignment and bow requirements for the bump bonding process. 

The introduction of more advanced packaging may solve many of these issues, allowing for the improvement of performance, yield and processing.
3D integration is a common widespread technology in industry, it allows tight packaging of sensor and readout chip. Furthermore it allows to stacks multiple chips in a single monolithic device. 
There are many technologies available for 3D integration, hybrid bonding is the most widely accepted: silicon covalent oxide bonding combined with copper diffusion bonding to provide connectivity between layers.
The process can be done for wafer to wafer (w2w) or die to wafer (d2w) assembly.
Through Silicon Vias (TSVs) allow multiple planes to be stacked and connected with external connections without the need of extensions or silicon interposers.
There are several advantages in using 3D integration in sensor to chip connection:
\begin{itemize}
\item Less space: 3D chips can be multi layer and do not need extension or interposers for external connections. Reduction of single layer thickness, after integration all supports can be removed
\item Very fine pitch bonding: down to a few micrometers
\item Better connections: faster and shorter than in circuit boards, with reduced dissipated power
\item Better performance: reduced input capacitance and lower noise
\item Layered design: e.g. sensor + analog electronics + digital electronics, each layer can be manufactured by different producers
\item Reduced thermal stress and increased heat dissipation since material is homogeneous, also increased robustness
\end{itemize}

Advanced packaging and wafer to wafer bonding would facilitate several applications both in HEP and outside of HEP: 4D tracking with high granularity LGADs, 3D integrated SiPM with advanced signal processing, chip tile array assembled at wafer level with no edges, small pixels connection (10 \rm{$\mu m$} or less) that reduces the input capacitance of thin sensors (e.g. thin LGADs) increasing both space and time precision, double sided connection for LGADs to have readout on both sides, stacked 3D integrated chip, 3D network of algorithm cells for advanced pattern recognition, zero mass trackers with very thin and highly populated layers, sensor stacks for high energy X-ray detection and decay detection. Finally it will advance the knowledge in substrate engineering, useful, for example, in the fabrication of buried p-n layers for LGADs, necessary for the production of, for example, DJ-LGAD and buried junction LGAD.

Therefore electronics and sensor advanced packaging provide a variety of technologies that can meet the needs of future particle physics experiments. 
Combining these capabilities with silicon technologies developed for HEP, such as LGADs and active edge sensors, will allow the design of sophisticated detector systems that can meet the increasing challenges of next generation experiments. 
The collaboration between research groups and industry with established expertise in advanced packaging is crucial for the successful introduction of this technology in the research community. 

Current availability of such technologies is mixed.  Leading edge foundries and nodes now include 3D processes (Intel Forveros, TSMC 3DFabric, Samsung x-cube) but are typically too expensive for HEP. Hybrid bonding is available from many vendors that already work with HEP.  The availability of TSVs is more limited. Small pitch TSVs are available from some foundries at advanced nodes ($>$32 nm), in Silicon-on-Insulator (SOI) wafers, and from specialty foundries.  Larger pitch TSVs inserted in a completed wafer (via-last) are available from a variety of packaging suppliers.  Reliable 3D devices with these large pitch ($>$20 $\mu m$) external TSV connections can be accessed now with resources available to HEP.  It is important to build design and fabrication experience with moderate scale companies now gearing up to address opportunities afforded by the US semiconductor initiative. 

It is urgent that this research is begun as it will take considerable effort and time in order to have the confidence required to use these techniques in future large-scale tracking systems.

\section{Mechanics}

Detector mechanics will play a significant role in future detectors' performance; the necessary improvements will require simulations, novel ways to reduce the total mass, as well as more integrated design concepts to save on material budgets and optimize performance. The increased segmentation has naturally lead to larger power densities requiring high performance material support structures with integrated services.  In many use-cases, the material in these structures can be the limiting factor for the tracking performance of the system.  Particle detectors at future colliders rely on ever more precise charged particle tracking devices, which are supported by structures manufactured from composite materials. Various engineering techniques able to solve challenges related to the design and manufacturing of future support structures have been developed.

Future particle colliders, such as the high luminosity Large Hadron Collider (HL-LHC), the Future Circular Collider (FCC-ee, FCC-hh), or the muon collider (MuC) will collide particles at unprecedented rates and present a harsh environment for future detectors. A holistic approach to the design and manufacturing of detector support structures will be necessary to achieve minimal weight systems and to potentially integrating electrical and cooling services into structures. Novel techniques, materials and other design and manufacturing solutions provide an avenue to solve challenges of increasingly more complex and large tracking detectors. Complex stresses in composite structures consisting of multiple parts are a consequence of different manufacturing techniques utilized (compression molding, oven-cured lamination, etc.).  These stresses pose a severe challenge to current simulation tools. Research and development efforts are underway to solve these challenges by exploring multi-functional composite structures manufactured from highly thermally conductive materials using different manufacturing techniques. Machine learning based topology optimization have been employed to fulfill the low-mass high-stiffness and thermal conductivity structure requirements.

Material savings due to novel approaches have the potential of reduction on the order of 30-50\% depending on more detailed R\&D studies. This is an ideal opportunity to explore the conjunction of latest techniques in composite engineering involving machine learning based algorithms for heat transfer, mechanical loading and micro-to-macro scale material response predictions for the performance of the support structure mechanics. Current R\&D efforts provide a headway for future detector support structures relying on transformational novel manufacturing techniques. 

Different structures are current under study to achieve these goals which include: carbon fiber support structures with integrated titanium pipes for $\rm{CO_2}$ cooling, etched silicon and peak cooling micro-channels, Kapton-based support structures and engineered air cooling. 

Historically, mechanics has had limited support in the US and international community. To reach the targeted performance of future tracking detector systems, an increase of resources will be required to be successful. 

\section{Collaborative development programs}

Collaborative efforts should be encouraged when possible to make efficient use of resources and reduce the financial burden on individual groups, and to enlarge the pool of expertise in the field. Especially critical is to encourage and support young researchers to engage and pursue intensive research programs to develop these new technologies, which will help to ensure an expert workforce is sustained for the long term. Considering the prohibitively high costs for productions of new technologies at commercial foundries, collaborative submissions can provide the only viable solution for research groups to access advanced production facilities. Such collaborations promise to significantly expand the possibilities for small experiments or small groups to join advanced detector R\&D programs, work with others in the field to build technology demonstrators, and help advance the solid-state tracking technology.


\bibliography{SnowmassBook-IF03} 
\bibliographystyle{ieeetr}


\end{document}